\documentclass[12pt,a4paper]{article}
\usepackage{amssymb} 
\begin{document}
\title{Momentum of cosmological acoustic field}
\author{Wojciech Czaja$^2$, Zdzis{\l}aw A. Golda$^{1,2}$, Andrzej Woszczyna\thanks{uowoszcz@cyf-kr.edu.pl}\,\,\,$^{1,2}$}
\date{}
\maketitle

\vspace{-10mm}

\begin{center}
{{$^1$} Astronomical Observatory, Jagiellonian University,\\  ul. Orla 171, 30--244 Krak\'ow, Poland}\\
\smallskip
{{$^2$} Copernicus Center for Interdisciplinary Studies,\\ ul. S{\l}awkowska 17, 31--016 Krak\'ow, Poland}
\end{center}
\bigskip
\bigskip 
\begin{abstract}
The {\it acoustic spacetime}  corresponding to
perturbed Friedman--Lema\^{\i}tre--Ro\-bert\-son--Walker universe
inherit the space isometries from the original FLRW model, 
but essentially differs in dynamics. The scale factor manifestly depends on the equation of state of the matter content. Despite the higher complexity of the background evolution the perturbation equation in this space is substantially simpler: 
the density perturbations obey d'Alembert equation. Canonical formalism reconstructed in the {\it acoustic spacetime\/} enables one to employ the Klein--Gordon scalar product. 
Consequently, the Fourier decomposition of the perturbation field provide the time-independent Fourier coefficients and the time-independent spectrum. 
The perturbation spectrum does not depend of the choice of the Cauchy hypersurface from which the data are collected. Noether constants associated with the six-parameter isometry group define the components 
of the momentum, hyperbolic momentum and angular momentum of sound.
\end{abstract}
\def\goa{{\mathfrak{a}}}
\def\gog{{\mathfrak{g}}}
\def\goi{{\mathfrak{i}}}
\def\gon{{\mathfrak{n}}}
\def\gox{{\mathfrak{x}}}
\def\sfx{{\sf x}}
\def\sfy{{\sf y}}
\def\sfz{{\sf z}}
\def\dd{p}
\def\DD{P}
\def\ddd{\underline{p}}
\def\DDD{\underline{P}}
\def\xx{\underline{x}}
\def\XX{\underline{X}}
\def\bfx{\mathbf x}
\def\bfxp{\mathbf x'}
\def\bfk{\mathbf k}
\def\bfn{\mathbf n}
\def\lap{\triangle}
\def\gRW{g}
\def\fchi{\chi_{k}}
\newcommand{\diag}{\mbox{diag}\,}
\def\pdf#1{\frac{\partial#1}{\partial\eta}}
\def\pds#1{\frac{\partial^2#1}{\partial\eta^2}}
\def\pdfz#1{\frac{\partial#1}{\partial\zeta}}
\def\pdsz#1{\frac{\partial^2#1}{\partial\zeta^2}}
\def\hang{\hangindent\parindent}
\def\textindent#1{\indent\llap{#1\enspace}\ignorespaces}
\def\litem{\par\hang\textindent}
\def\subitem{\par\indent \hangindent2\parindent\textindent}
\def\peta{\partial_\eta}
\def\petaeta{\partial_{\{\eta,2\}}}
\def\pochodna{\partial}
\def\ped{k}
\def\k0{K}
\def\k1{K_r}
\def\k2{K_l}
\def\mmu{\mathcal{C}}
\def\llambda{\triangle\mathcal{E}}
\def\uu{u_k(x^{\mu},n^{i})}
\def\uuast{u^{\ast}_k(x^{\mu},n^{i})}
\def\uk{u_{k}}
\def\ukast{u^{\ast}_{k}}
\def\ukprim{u_{k'}}
\def\ak{a_{k}}
\def\akast{a^{\ast}_{k}}
\def\akprim{a_{k}'}
\def\gg{\gamma(x^{\mu},n^{i})}
\def\varSigma{\mathit\Sigma}

\section{Introduction}
\label{sec:01}

Cosmological acoustic field originates due to small perturbation of the Robertson--Walker background and contributes to the microwave background temperature fluctuations. The perturbation formalisms describing this phenomenon, initiated by Lifshitz~\cite{1}, Lifshitz and Khalatnikov~\cite{2}, Landau and Lifshitz~\cite{3} and continued by many different authors and groups, can be divided into four categories:

\litem{1.} Formalisms which adopt gauge-invariant combinations of metric potentials, metric components and their first derivatives. The propagation equations for them are easy to derive Bardeen~\cite{4},  Brandenberger, Kahn and Press~\cite{5}, but  the interpretation of these variables is not straightforward Bardeen~\cite{4}, Katz et al.~\cite{6}, Deruelle et al.~\cite{7}, and  Deruelle and~Uzan~\cite{8}.

\litem{2.} {Gauge-specific formalisms}.\footnote{Cosmological literature does not precisely distinguish between the {\em gauge-invariant\/} and {\em gauge-specific\/} approaches. There is a tendency to give the name of {\em gauge-invariant\/} to all these theories that eliminate the pure gauge perturbations, even if they do eliminate the gauge freedom itself.
The expressions such as ``gauge invariant variables like the curvature perturbation on a constant Hubble surface'' are often met and rarely criticised.} Particularly these, which appeal to the orthogonal gauge~\cite{4,9}, the zero shear gauge~\cite{4}, or the longitudinal gauge~\cite{10}.

\litem{3.} The gauge invariant formalisms operating with 
the hydrodynamic quantities like shear, divergence of the fluid flow, density gradients or Laplacian~\cite{11,12,13,14,15}. These quantities are local, and as such, obviously are gauge-invariant. Their geometrical interpretation is clear but it is difficult to construct the propagation equations in the partial differential form.

\litem{4.} The formalisms operating with gauge invariant variables of the desired dynamical properties~\cite{16,17,18,19}. The theories have Hamiltonian form. Perturbations are identified with the scalar fields of  time-dependent mass. No particular geometrical interpretation is developed.
\medskip 

More about the mutual relationships between different formalisms can be found in~Brechet et al.~\cite{20}. Each category gives insight into different aspect of the perturbation problem. Non of them is known as complete and commonly accepted theory of cosmological density perturbations. 

The theory of sound in non static media has developed in roughly the same period. Research refers both to theoretical~\cite{21,22,23,24,25,26,27} and engineering~\cite{28,29,30,31,32}  aspects. The key feature that distinguishes the acoustics of fluids in motion is the absence of the conserved quantities: energy, momentum and angular momentum~\cite{33} of sound. Non-conservation of the acoustic energy or momentum leads to the idea of the energy-momentum interchange between the wave and the background flow. The more radical formulation states that no momentum at all can be assigned to sound wave propagating in non-static environment (see McIntyre~\cite{33} critical review). 

{\it Acoustic geometry\/}~\cite{34,35,36} provides the good insight into the acoustic energy-momentum problem and points out the essential difficulties~\cite{37}. For arbitrary fluid flow the Unruh metric~\cite{34,35,36} may not admit isometries, and consequently, the energy or momentum cannot be introduced as the Noether constants.

The acoustics of non static media has several tangent points with the cosmological perturbations theory prior the recombination epoch. In both cases: 

\litem{a)} the environment evolves, and the energy is not conserved,
\litem{b)} the density perturbations propagate as waves,
\litem{c)} inhomogeneities are small, the linear theory is adequate,
\litem{d)} splitting the solutions into the background component and the perturbation is not unique (gauge problem).

\noindent
Despite these parallels the theory of sound propagation in fluid flows remains unknown for cosmologists. To our best knowledge the questions {\it what is the momentum of the sound wave\/}~\cite{38} has never been asked in cosmology.

\medskip
An attempt to answer this question is the aim of this paper. The problem is simpler than in classical acoustics, because the Robertson--Walker spacetime admits the six parameter group of isometries~\cite{39}. Killing vectors are spacelike. The resulting Noether constants can be easily identified with the momentum and angular momentum components.
Following~Unruh~\cite{34,35}, Visser~\cite{36}, Golda and Woszczyna~\cite{40,41,42} we construct the acoustic spacetime in which the perturbations behave as the scalar field (the propagation equations have the d'Alembert equation form). Then, we introduce the energy momentum tensor for such field and, following classical methods~\cite{6,7,8,43}, we construct the conserved currents and constants of motion.

We concentrate on adiabatic perturbations. We do not specify any particular equation of state, we claim however that $p=p(\epsilon)>0$ and $\delta p/\delta\epsilon>0$ (the speed of sound $c_{\rm s}$ is positive). We suppose that some difficulties to express the behaviour of multi component fluids in the framework of general relativity (with the famous ``frozen amplitudes'' problem included) may come from the insufficiently precise formulation of the expanding fluid acoustics. The single fluid case is good for study the issues.

We adopt the synchronous system of reference, where the gauge freedom is limited by the constraint $\delta g_{0\mu}=0$. This constraint is a good compromise between full gauge freedom and completely fixed gauge theory. Computations in synchronous systems are substantially easier, while the residual gauge freedom still admits changes of the constant time hypersurfaces. The gauge-invariance evoked below means invariance against this residual freedom. The freedom is enough to introduce the hypersurface-independent variables and the time-independent perturbation spectrum.\footnote{We are sure that the  $\delta g_{0\mu}=0$  constraint can be relaxed with no harm to the presented results.}

This paper consists of 8~sections. The short section~\ref{sec:02} revokes the definition of scalar perturbations on the Robertson--Walker background. In the section~\ref{sec:03} we derive the partial-differential analogue of the Lifshitz--Khalatnikov equations. The gauge-invariant variables are introduced in section~\ref{sec:04}. Propagation equation for the new variables is reduced to the wave equation form (section~\ref{sec:05}). Construction of the acoustic spacetime for the cosmological density perturbations is given in section~\ref{sec:06}. We reduce the propagation equation to the d'Alembert equation in this space.  Isometries, the momentum of acoustic field and its spectrum are discussed the section~\ref{sec:07}. Some concluding remarks are given in section~\ref{sec:08}. A short code written in {\it Mathematica\/} is attached in the Appendix (section~\ref{sec:09}).

Throughout this paper we use the convention $8\pi G=1$, $c=1$. The following notation is adopted:\\ 
$a(\eta)$ --- conformal scale factor,\\
$\eta$ --- conformal time,\\ 
$x^k = \{\sfx, \sfy, \sfz\}$ --- Cartesian 3-space coordinates,\\
$x^\mu =\{\eta, \sfx, \sfy, \sfz\}$ --- Cartesian spacetime coordinates,\\
$^{(3)}\!g_{mn}$ --- the metric of the maximally symmetric 3-dimensional space. The curvature~$K$ is an arbitrary real number.
 
Where the confusion is unlikely, we abbreviate the notation by writing $\bfx$ for $x^\mu$ and $\sfx$ for $x^k$. Dot in $\sfx{\cdot}\sfx$ stands for $\sfx^2+\sfy^2+\sfz^2$.   We never explicitly express the time or space dependence of the metric tensor $g$ (we hope, this is obvious), but we always keep this dependence in the metric corrections $\mathcal{C}(\bfx)$ and $\mathcal{E}(\bfx)$. Vertical lines as in $\mathcal{E}(\bfx)_{|i}^{\phantom{|i}|j}$ mean the covariant differentiation in the space $^{(3){\!}}g_{mn}$. The same $\mathcal{E}(\bfx)_{|i}^{\phantom{|i}|j}$ is later abbreviated to $\lap\,\mathcal{E}(\bfx)$.

\section{Scalar perturbations}
\label{sec:02}

Consider the Robertson--Walker spacetime as the ground state,
		\begin{equation}
\gRW_{\mu\nu}=
a^2(\eta)
\left(\!\!
\begin {array} {cc}
 - 1 & 0\\
 0 &{}^{(3){\!}}g_{mn}
 \end {array}
\!\!\right),
		\label{00}
		\end{equation}
where ${}^{(3){\!}}g_{mn}$ stands for the maximally symmetric 3-dimensional space metric tensor, expressed in Cartesian coordinates
		\begin{equation}
{}^{(3){\!}}g_{mn}=
\left(
1+\frac14 K\, \sfx{\cdot}\sfx 
\right)^{-2}\delta_{mn}. 
		\label{01}
		\end{equation}
We introduce the small correction $\delta g_{\mu \nu}$		
		\begin{equation}
\gRW_{\mu\nu}\longrightarrow\gRW_{\mu\nu}+\delta g_{\mu\nu}
		\label{02}
		\end{equation}
demanding that
		\begin{eqnarray}
			\label{poprawka01}
\delta g_{m n}&=&\nabla_{\!m}\nabla_{\!n}\, \mathcal{E}(\bfx)+\frac{1}{3}\left( \mathcal{C}(\bfx)-\triangle\, \mathcal{E}(\bfx)\right)g_{m n},\\
\delta g_{\mu 0}&=&0. 
		\label{poprawka02}
		\end{eqnarray}
$\nabla_{\!m}$ and $\triangle$ denote the covariant derivative and the Laplacian, respectively --- both are calculated in the static space ${}^{(3){\!}}g_{mn}$. $\mathcal{C}(\bfx)$ and $\mathcal{E}(\bfx)$ are arbitrary functions dependent on space and time. The correction (\ref{poprawka01}) and (\ref{poprawka02}) is the most general formula for the scalar perturbations~\cite{44} in the synchronous system of reference.

\section{Propagation equations for metric potentials}
\label{sec:03}

The Einstein equations
		\begin{equation}
G_{\mu\nu}=T_{\mu\nu}
		\label{rownania_Einsteina}
		\end{equation}
with the hydrodynamic energy momentum tensor		
		\begin{equation}
T^{\mu}_{\phantom{\mu}\nu}=\diag(-\epsilon,p(\epsilon),p(\epsilon),p(\epsilon))
		\label{tensor_energii_pedu}
		\end{equation}
are expanded to the first perturbation order with respect to the metrics corrections. The zero-order equations reduce to the Friedman equations
\begin{eqnarray}
			\label{Friedman01}
3\frac{{a'}^2(\eta)}{a^4(\eta)}+3\frac{K}{a^2(\eta)}&=&\epsilon(\eta),\\
-2\frac{{a''}(\eta)}{a^3(\eta)}+\frac{{a'}^2(\eta)}{a^4(\eta)}-\frac{K}{a^2(\eta)}&=&p(\eta),
		\label{Friedman02}
		\end{eqnarray}
which define the dynamics of the background spacetime. In the first perturbation order one obtains the Lifshitz--Khalatnikov equations		
		\begin{eqnarray}
		\label{RL01}
h_i^{\phantom{i}l|j}{}_{|l}+
h^j_{\phantom{j}l|i}{}^{|l}-
h_i^{\phantom{i}j|l}{}_{|l}-
h_{|i}^{\phantom{|i}|j}+\pds{h_i^{\phantom{i}j}}
+2a\,H\pdf{h_i{}^j}-4Kh_i^{\phantom{i}j}=0,~\mbox{for}~i\neq j,\\
\pds{h}+(2+3c_{\rm s}^2)a\, H\pdf{h}+\frac12(1+3c_{\rm s}^2)(h_l{}^{m|l}{}_{|m}-h_{|l}{}^{|l}-2Kh)=0.
		\label{RL02}
		\end{eqnarray}
where $h$ is the abbreviation for the metric perturbation:  
$h_{mn}=\delta g_{mn}$, 
$h=g^{mn}h_{mn}$, 
the stroke stands for the covariant space-derivative $\nabla_{\!n}$  ($T_{|n}=\nabla_{\!n} T$) and $c_{\rm s}$ stands for the sound velocity: $c_{\rm s}^2=p'(\eta)/\epsilon'(\eta)$. $H=a'(\eta)/a^2(\eta)$ is the Hubble parameter. In this notation the scalar perturbations (\ref{poprawka01})--(\ref{poprawka02}) read
		\begin{eqnarray}
			\label{ta_sama_poprawka01}
h_i{}^j&=&\frac13(\mathcal{C}(\bfx)-\mathcal{E}(\bfx)_{|l}{}^{|l})\delta_i{}^j+
\mathcal{E}(\bfx)_{|i}{}^{|j},\\
h&=&h_i{}^i=\mathcal{C}(\bfx)
		\label{ta_sama_poprawka02}
		\end{eqnarray}
Inserting (\ref{ta_sama_poprawka01}) and (\ref{ta_sama_poprawka02}) into
 the equations (\ref{RL01}) and (\ref{RL02}) one obtains the propagation equations for the metric potentials $\mathcal{C}$ and $\mathcal{E}$ in their partial differential form
		\begin{eqnarray}
			\label{RLCE01}
\pds{}\mathcal{E}_{|m}{}^{|n}+2a\, H\pdf{}\mathcal{E}_{|m}{}^{|n}+
\mathcal{E}_{|ipml}\,g^{in}g^{pl}-\frac23\mathcal{E}_{|iplm}\,g^{ip}g^{ln}\nonumber\\
{}+\mathcal{E}(\bfx)_{|impl}(g^{il}g^{np}-g^{in}g^{pl})-
\frac13\mathcal{C}_{|m}{}^{|n}-4K \mathcal{E}_{|m}{}^{|n}=0,~
\mbox{for}~m\neq n,\\
			\label{RLCE02}
\pds{}\mathcal{C}+(2+3c_{\rm s}^2)a\, H\pdf{}\mathcal{C}+(1+3c_{\rm s}^2)\nonumber\\
{}\times\left[\frac12\mathcal{E}_{|lhsm}(g^{lm}g^{hs}-\frac13g^{lh}g^{sm})-
\frac13\left(\mathcal{C}_{|r}{}^{|r}+3K \mathcal{C}\right)\right]=0.
		\end{eqnarray}
By adopting the identity
		\begin{eqnarray}
			\label{RLT1}
\mathcal{E}_{|ipml}(g^{in}g^{pl}-g^{ip}g^{nl})+
\mathcal{E}_{|impl}(g^{il}g^{np}-g^{in}g^{pl})
-4K\mathcal{E}_{|m}{}^{|n}=0,
		\end{eqnarray}
one can reduce the first equation (\ref{RLCE01}) to
		\begin{eqnarray}
			\label{drugiego_rzedu}
\left[\pds{}\mathcal{E}+2a\, H\pdf{}\mathcal{E}+\frac13\left(\mathcal{E}_{|l}{}^{|l}-
\mathcal{C}\right)\right]_{|m}^{\phantom{|m}|n}=
0,\quad \mbox{for}~m\neq n.
		\end{eqnarray}
To fulfil the equality $f(x^\mu)_{|m}{}^{|n} =0$, $\forall m \neq n$ in arbitrary coordinate system, the function $f(x^\mu)$ must be a constant. Then, $f$ obviously satisfies $f(x^\mu)_{|m}{}^{|m}=0$ which implies also that
		\begin{eqnarray}
			\label{RL1F}
\left[\pds{}\mathcal{E}+2a\, H\pdf{}\mathcal{E}+\frac13\left(\mathcal{E}_{|l}{}^{|l}-
\mathcal{C}\right)\right]_{|m}^{\phantom{|m}|m}=0.
		\end{eqnarray}
On strength of the identity
		\begin{eqnarray}
			\label{RLT2}
\mathcal{E}_{|lhsm}\left(g^{lm}g^{sh}-g^{lh}g^{sm}\right)-2K\mathcal{E}_{|m}{}^{|m}=0,
		\end{eqnarray}
the second equation (\ref{RLCE02}) simplifies to 
		\begin{eqnarray}
			\label{RL2F}
\pds{}\mathcal{C}+(2+3c_{\rm s}^2)a H\pdf{}\mathcal{C}+\frac13(1+3c_{\rm s}^2)\!
\left[\!\left[\mathcal{E}_{|l}{}^{|l}-\mathcal{C}\right]_{|m}\!\!{}^{|m}+3K\!\!\left[\mathcal{E}_{|l}^{\phantom{|l}|l}-\mathcal{C}\right]\!\right]=0.
		\end{eqnarray}
Finally, the propagation equations take the form 
		\begin{eqnarray}
			\label{LKh01}
\frac{\partial^2\llambda}{\partial\eta^2}&=&-2\frac{a'}{a}\frac{\partial\llambda}{\partial\eta}-
\frac{1}{3} \triangle\left(\llambda-\mmu\right),\\
\frac{\partial^2\mmu}{\partial\eta^2}&=&-(3c_{\rm s}^2+2)\frac{a'}{a}\frac{\partial\mmu}{\partial\eta}
-\left(\frac13+c_{\rm s}^2\right)
\left(\triangle+3K \right)
\left(\llambda-\mmu\right),
			\label{LKh02}
		\end{eqnarray}
These equations are the partial differential analogue to the Lifshitz--Khalatnikov system.  Neither the metric potentials nor the density contrast evaluated from them 
		\begin{eqnarray}
\delta\epsilon(\bfx)=\frac{1}{3a^2}\left[3\frac{a'}{a}\pdf{\mmu}+
(\triangle+3K)(\llambda-\mmu\right]. 
			\label{delta_e}
		\end{eqnarray}
are observables. All of them are ambiguous due to the existing gauge freedom.

\section{Gauge-independent variables}
\label{sec:04}

When the gauge is fixed, like the fluid orthogonal gauge or the zero shear gauge, all the quantities are {\em gauge-specific}. There is no standard technique to introduce gauge-invariant quantities on fixed hypersurfaces, neither to compare gauge-specific quantities between different gauges. For the properly constructed gauge-invariant spectrum the existence of gauge freedom is indispensable.

In the case we discuss in this paper, the gauge freedom is guaranteed by two arbitrary space dependent functions $\mathcal{G}_1(x^k)$ and $\mathcal{G}_2(x^k)$. The gauge solutions for $\llambda$ and $\mmu$ are
		\begin{eqnarray}
			\label{LMgauge01}
\llambda_g(x^\mu)&=&-\mathcal{G}_1(x^k)-2\int\!\!\frac{1}{a(\eta)}{\rm d}\eta\,\,\triangle\mathcal{G}_2(x^k),\\
\mmu_g(x^\mu)&=&\mathcal{G}_1(x^k)+6\frac{a'(\eta)}{a^2(\eta)}\mathcal{G}_2(x^k)+2\int\!\!\frac{1}{a(\eta)}{\rm d}\eta\,\,\triangle\mathcal{G}_2(x^k). 
			\label{LMgauge02}
		\end{eqnarray}
Not eliminating this freedom, we look for quantities of physical interest, which are independent of $\mathcal{G}_1(x^k)$ and $\mathcal{G}_2(x^k)$. 

Below we abbreviate the notation by writing $\bfx$ for $x^\mu$ and $\sfx$ for $x^k$. 
Our construction of the gauge-independent variables is based on the fact that the pure-gauge energy density perturbation
		\begin{eqnarray}
\delta\epsilon_g(\bfx)=\mathcal{G}_2({\sfx})\frac{\epsilon'(\eta)}{a(\eta)}=
\mathcal{G}_2({\sfx})\dot{\epsilon}(t)
			\label{density_gauge}
		\end{eqnarray}
forms the product of $\mathcal{G}_2(\sfx)$ and the time derivative of the background energy density $\dot{\epsilon}(t)$ (the formula ${\rm d}\eta=a(t){\rm d}t$ connects the two different time parameters). The formula (\ref{density_gauge}) can be easily obtained by inserting the system (\ref{LMgauge01}) and (\ref{LMgauge02}) into (\ref{delta_e}). Fictitious perturbations for other scalar quantities (the expansion rate, the scalar curvature, etc. ) have the same structure
		\begin{eqnarray}
\left[\!\frac{\delta\epsilon(\bfx)}{\dot{\epsilon}(t)}\!\right]_g=
\mathcal{G}_2({\sfx}),~\left[\!\frac{\delta\theta(\bfx)}{\dot{\theta}(t)}\!\right]_g=
\mathcal{G}_2({\sfx}),~\left[\!\frac{\delta R(\bfx)}{\dot{R}(t)}\!\right]_g=
\mathcal{G}_2({\sfx}),~\mbox{etc. }
			\label{aaa_gauge}
		\end{eqnarray}
Differentiation with respect to time removes the gauge dependence $\mathcal{G}_2({\sfx})$. Hence, a~simple mnemotechnic rule arises, that allows one to convert any fractional perturbation of scalar quantity (the so called {\it contrast\/}) into the corresponding dimensionless gauge-independent variable
		\begin{eqnarray}
\left[\!\frac{\delta\epsilon(\bfx)}{{\epsilon}(t)}\!\right]
\longrightarrow \left[\!\frac{\delta\epsilon(\bfx)}{{\dot{\epsilon}(t)}}\!\right]^{\mbox{. }}\!\!,
~\left[\!\frac{\delta\theta(\bfx)}{{\theta}(t)}\!\right] 
\longrightarrow \left[\!\frac{\delta\theta(\bfx)}{\dot{\theta}(t)}\!\right]^{\mbox{. }}\!\!,~
\left[\!\frac{\delta R(\bfx)}{{R}(t)}\!\right] 
\longrightarrow \left[\!\frac{\delta R(\bfx)}{\dot{R}(t)}\!\right]^{\mbox{. }}\!\!,~\mbox{etc. }
			\label{gi}
		\end{eqnarray}
The rule consists in completing the expression for {\it contrast} with two derivative dots: one in the denominator, and one over the whole fraction.

\section{Propagation of sound}
\label{sec:05}

Arbitrary time dependent factor $f(t)$ multiplying the perturbation amplitude $Y(\bfx)$ 
		\begin{eqnarray}
Y(\bfx) \longrightarrow f(t)Y(\bfx)		
			\label{scalling}
		\end{eqnarray}
does not affect the gauge-invariance. The formulae (\ref{gi}) define the families of gauge independent quantities, with the amplitude scaling freedom (\ref{scalling}). The choice of the factor~$f(t)$ is one of the possible tools to find the {\em canonical form\/} for the propagation equation. We guess the factor $f$ in the form $f(t)=\dot{a}^2(t)$, and introduce 
		\begin{eqnarray}
\gamma(\bfx)=\dot{a}^2(t)\left[\!\frac{\delta\epsilon(\bfx)}{\dot{\epsilon}(t)}\!\right]^{\mbox{. }}
			\label{gamma}
		\end{eqnarray}
to define the gauge-invariant variable for the energy density perturbation. Returning to the conformal time $\eta$ we write
		\begin{eqnarray}
\pds{\gamma(\bfx)}+
\left[2\frac{\goa'(\eta)}{\goa(\eta)}-\frac{c_{\rm s}'(\eta)}{c_{\rm s}(\eta)}
\right]\pdf{\gamma(\bfx)}-c_{\rm s}^2(\eta)\triangle \gamma(\bfx)=0,
			\label{variable}
		\end{eqnarray}
where the function $\goa(\eta)$ reads
		\begin{eqnarray}
\goa(\eta)=a(\eta)\sqrt{\frac{p(\eta)+\epsilon(\eta)}{3c_{\rm s}(\eta)H^2(\eta)}}. 
			\label{apisane}
		\end{eqnarray}
The background dependent quantities: Hubble parameter $H$, pressure $p$, the energy density $\epsilon$ and the sound velocity $c_{\rm s}$ involved in (\ref{apisane}) are uniquely determined by Friedman equations and the equation of state. 

The equation (\ref{variable}) describes the acoustic waves propagating in the expanding environment with variable sound velocity $c_{\rm s}(\eta)$. 

Proof of the equation (\ref{variable}) consists in simultaneous computer aided reduction of the system of equations: 
(\ref{Friedman01}), (\ref{Friedman02}),
(\ref{LKh01}), (\ref{LKh02}), (\ref{delta_e}), (\ref{gamma}), (\ref{variable}). 
A short code written in Mathematica is presented in the Appendix. We use {\it normal forms\/} of commands and expressions, where possible. The full code for sections \ref{sec:02}--\ref{sec:05} written in Tensorial package is available from~\cite{45}.

\section{Acoustic spacetime}
\label{sec:06}

Appropriate redefinition of the time parameter completes the construction of the acoustic spacetime~\cite{34,35,36}. To avoid the explicit time-dependence of the sound velocity $c_{\rm s}(\eta)$ in the equation (\ref{variable}) we introduce 
		\begin{eqnarray}
\zeta=\int^\zeta\!\!c_{\rm s}(\eta)\,{\rm d}\eta. 
			\label{dzeta}
		\end{eqnarray}
The resulting propagation equation 
		\begin{equation}
\pdsz{\gamma(\gox)}+
2\frac{\goa'(\zeta)}{\goa(\zeta)}
\pdfz{\gamma(\gox)}-\triangle \gamma(\gox)
=0,
			\label{jawna}
		\end{equation}
is the explicit form of the d'Alembert equation
		\begin{equation}
\gog_{\mu\nu}\nabla^{\mu}\nabla^{\nu}{\gamma(\gox)}=0,
			\label{dAlembert}
		\end{equation} 
in the spacetime $\gox= \{\zeta,\sfx,\sfy,\sfz\}$ with the metric 
		\begin{equation}
			\gog_{\mu\nu}=\goa^2(\zeta)\left(\!\!
			\begin{array}{cc}
 				- 1 & 0\\
 				0 &{}^{(3){\!}}g_{mn}
 			\end{array}
			\!\!\right),~\mbox{where}~\goa(\eta(\zeta))=a\sqrt{\frac{p+\epsilon}{3c_{\rm s}H^2}}. 
		\label{Mprim}
		\end{equation}
Equations (\ref{jawna}) and(\ref{dAlembert}) say that the sound in the Robertson--Walker spacetime $\cal M$ with the scale factor $a(\eta)$, propagate as massless scalar field (\ref{dAlembert}) in another Robertson--Walker spacetime $\cal M'$ (\ref{Mprim}). 
Sachs and Wolfe~\cite{16} reported wave-like behaviour of scalar perturbations in the flat, radiation-dominated universe. The equations (\ref{dAlembert}) and (\ref{Mprim}) may be considered as the generalization of the Sachs--Wolfe theorem to spatially curved Robertson--Walker models and to arbitrary equation of state of the barotropic form $p=p(\epsilon$). 

In the following sections we drop the sign prime. The coordinates $\{\zeta,\sfx,\sfy,\sfz\}$  we briefly denote by the same letter $\bfx$.

\section{Symmetries, momenta and spectrum of sound}
\label{sec:07}

\def\OT{T}
\def\OM{M}
\def\OL{L}
\def\KT{K_{\OT}}
\def\KM{K_{\OM}}
\def\KL{K_{\OL}}

\def\ik{{\int\!\! {\rm d}k}}
\def\ikk{{\int\!\!{\rm d}k \int\!\! {\rm d}k '}}

\def\ak{a_{k}}
\def\akast{a^{\ast}_{k}}

\def\akprim{a_{k'}}
\def\akprimast{a^{\ast}_{k'}}
\def\ukprimast{u^{\ast}_{k'}}

\def\chik{\chi_{k}}
\def\chikast{\chi^{\ast}_{k}}

We introduce Lagrangian for the acoustic field
		\begin{equation}
{\cal L}=-\frac12\gog_{\rho\sigma}\pochodna^{\rho}{\gamma(\bfx)}\,\pochodna^{\sigma}{\gamma(\bfx)}
			\label{lagrangian}
		\end{equation}
and derive the canonical energy-momentum tensor
		\begin{equation}
{\cal T}^{\mu\nu}=\pochodna^{\mu}{\gamma(\bfx)}\,\pochodna^{\nu}{\gamma(\bfx)}-
\frac12 \gog^{\mu\nu} \gog_{\rho\sigma}\pochodna^{\rho}{\gamma(\bfx)}\,\pochodna^{\sigma}{\gamma(\bfx)}. 
			\label{akousticTEM}
		\end{equation} 
The propagation equation (\ref{dAlembert}) is the Lagrange equation for $\cal L$, or equivalently, it follows from the $\nabla_{\!\nu}{\cal T}^{\mu\nu}=0$ identity. Both ${\cal L}$ and ${\cal T}^{\mu\nu}$ are gauge-invariant, therefore, all the conserved quantities constructed by means of them have the same property. The procedure below follows that of  Katz et al.~\cite{6}, Deruelle et al.~\cite{7}, and  Deruelle and~Uzan~\cite{8} except for the background space is (\ref{Mprim}) and the field $\gamma(\bfx)$ is already gauge-invariant. 
We consider conserved currents ${\cal J}^{(i)}_{\mu}={\cal T}_{\mu\nu} K^{(i)\nu}$.
The six space-like isometries $K^{(i)}$ immediately provide the six Noether integrals $\pi(K^{(i)})$ 
		\begin{equation}
\pi(K^{(i)})
=\int_{\varSigma}{\cal J}^{(i)}_{\mu}{\rm d}\varSigma^\mu 
=\int_{\varSigma}\!\! {\cal T}_{\mu\nu} K^{(i)\nu}\,{\rm d}\varSigma^\mu = \mbox{const}^{(i)} \label{pedy1}
		\end{equation}
where $\varSigma^\mu$ is an arbitrary Cauchy surface. 
The algebra of Killing vectors, and consequently, the interpretation of constants (\ref{pedy1}) depend on the sign of curvature. For Killing basis chosen as $K^{(i)}=\{\KT^{(i)},\KL^{(i)}\}$ 
		\begin{eqnarray}
K_T^{(i)j}&=&\delta^{ij}-\frac{K}{4} \left[\sfx{\cdot}\sfx\, \delta^{ij}-2x^i x^j-\sqrt{K}\sum_{k=1}^{3}\epsilon^{ijk} x^k\right]
		\label{eq:killingT}\\
K_L^{(i)j}&=&\sum_{k=1}^{3}\epsilon^{ijk} x^k
		\label{eq:killingL}\\
K_L^{(i)0}&=&K_T^{(i)0}=0
		\end{eqnarray}
generators of infinitesimal isometries  
		\begin{equation}
\OT^{(i)}=-\goi \,\KT^{(i)j}\partial_j,\hspace{3mm}  \OL^{(i)}=-\goi\, \KL^{(i)j}\partial_j \label{generatory}
		\end{equation}
satisfy the commutation relations
		\begin{eqnarray}
\left[\OT^{(i)},\OT^{(j)}\right]&=&\goi \sqrt{4K} \sum_{k=1}^{3}\epsilon^{ijk} \OT^{(k)}\label{komutator1}\\
\left[\OL^{(i)},\OL^{(j)}\right]&=&-\goi   \sum_{k=1}^{3}\epsilon^{ijk} \OL^{(k)}\label{komutator2}\\
\left[\OL^{(i)},\OT^{(j)}\right]&=&-\goi   \sum_{k=1}^{3}\epsilon^{ijk} \OT^{(k)}\label{komutator3}
		\end{eqnarray}
For $K>0$, both $\KT^{(i)}$ and $\KL^{(i)}$ assure the angular momenta conservation (six constants of motion). 
For $K=0$ the right-hand side of (\ref{komutator1}) vanishes and operators $\OT^{(i)}$ generate translations. Translations  $\KT^{(i)}$ conserve the momentum (3 constants of motion), while the rotations  $\KL^{(i)}$ conserve the angular momentum  (the next 3 constants of motion). For $K<0$ vectors $\KT^{(i)}$ correspond to hyperbolic momentum~\cite{46,47,48}. The absence of the time isometry\footnote{Conformal Killing vectors in both spaces $\cal M$ and $\cal M'$ are not identical. The conserved quantities related to conformal isometries~\cite{6,7,8} of the space~$\cal M'$ are a~separate issue, not discussed in this paper.} breaks down the energy conservation. 

The Casimir operator 
		\begin{equation}
\sum_{i=1}^{3}\OT^{(i)} \OT^{(i)}
=-\Delta\label{Casimir}		
		\end{equation}
is equal to minus Laplacian. $\OT^{(i)}$ and $\Delta$ commutate
		\begin{equation}
\left[T^{(i)},\Delta\right]= 0\label{komutator3c}
		\end{equation}
hence, each pair $\{\OT^{(i)}, \Delta\}$, $i=1,\ldots,3$  has common eigenfunctions.
There are no common eigenfunctions for pairs $\{\OT^{(i)}, \OT^{(j)}\}$ with $i\neq j$ and $K\neq 0$. Solutions $\uu$ to equation (\ref{dAlembert}) which lie in the kernel of operator $n_{i}\OL^{(i)}$, and are eigenfunctions of the operator $n_{i}\OT^{(i)}$ 
		\begin{eqnarray}
&n_{i}\OL^{(i)} \uu&=0\\
&n_{i}\OT^{(i)} \uu&=(\ped-\goi \sqrt{-K})\uu 
	\label{wlasnepedu}
		\end{eqnarray}
define {\it plane waves\/} of the wavenumber $k$, propagating in the direction $n^{i}$.
On the strength of (\ref{komutator3c}) the same functions are also the eigenfunctions of the Laplace operator
		\begin{equation}
\Delta \uu = -(\ped^2-K)\uu
	 \label{helmholtz}		
		\end{equation}
Since $\Delta$ is Hermitian, the functions $\uu$ of different $k$ are orthogonal on constant time hypersurfaces. When separated they read
		\begin{equation}
u_k=\frac{1}{\goa(\zeta)}\fchi(\zeta)F(\sfx,\gon,k). \label{product}
		\end{equation}
The time dependent factor $\fchi(\zeta)$ obey the evolution equation
		\begin{eqnarray}
K-k^2&=&\frac{\fchi''(\zeta)}{\fchi(\zeta)}+2\frac{\goa'(\zeta)}{\goa(\zeta)}\frac{\fchi'(\zeta)}{\fchi(\zeta)}. 
	\label{czasowe}		
		\end{eqnarray}
and implicitly depend on the equation of state. The space-dependent solutions $F(\sfx,\gon,k)$ to the Helmholtz equation (\ref{helmholtz}) are 
		\begin{equation}
F(\sfx,\gon,k)= \frac{1}{(2\pi)^3/2}\left(\frac{-\sqrt{-K}\,\gon{\cdot} \sfx+1-\frac{K}{4}\sfx{\cdot} \sfx}{1+\frac{K}{4}\sfx{\cdot} \sfx}
\right)^{-1+\frac{\goi\,\ped}{\sqrt{-K}}}.\label{Shapiro}		
		\end{equation}
In the $K=0$ limit $F(\sfx,\gon,k)$ tend to $e^{\goi\, k(\gon{\cdot} \sfx)}$. For that case the spectral analysis of the acoustic field resembles that for electromagnetic field~\cite{49,50,51}. 
For the negative space curvature functions $F(\sfx,\gon,k)$ are Shapiro functions~\cite{52}. {\em Principal series\/} characterised by positive wavenumbers $k>0$ consists of functions orthogonal and complete in $L^2$~\cite{53,54}.
{\em Supplementary series\/} of regular, bounded, non-oscillating and non-orthogonal functions $F(\sfx,\gon,k)$ with imaginary wavenumber 
$k\in(0, \sqrt{K})$ are redundant to expand the square integrable perturbations. However, they contribute to Fourier decomposition of weakly homogeneous stochastic processes~\cite{55,56}. A single function of supplementary series is also needed for completeness in the quantum theory of charge~\cite{57}. 
In both cases, flat $(K=0)$ or open $(K<0)$ universes, the Fourier bases are denumerable, and the spectrum is continuous. 

For positive curvature the functions $F(\sfx,\gon,k)$ are known as Sherman--Volobuyev functions~\cite{58,59}. The spectrum of Beltrami--Laplace operator is numerable. The Sherman--Volobuyev functions do not form the complete set, neither are orthogonal~\cite{60}, and therefore, the case of closed universe requires a separate treatment. 

In what follows, we limit ourselves to non-positive curvature and to positive wave numbers (continuos spectrum and principal series). We consider the acoustic wave $\gg$ propagating in the direction $n^i$, i. e. an arbitrary solution to the equation (\ref{dAlembert}) that can be expanded
		\begin{equation}
\gg=\int_{0}^{\infty}\! (\ak \uu+\akast \uuast) {\rm d}k \label{dekompozycja}
		\end{equation}
in orthogonal basis $\uu$  
		\begin{equation}
(\uk,\ukprim)_{\cal KG}=\frac{k}{k-\goi\sqrt{-K}}\delta(k -k ') \label{norma}
		\end{equation}
with Fourier coefficients
		\begin{equation}
\ak=\left(1-{\goi\over{k}}\sqrt{-K}\right)(\uk,\gamma)_{\cal KG}. \label{wspolczynniki1}
		\end{equation}
Symbol $( , )_{\cal KG}$ means the Klein--Gordon scalar product
		\begin{equation}
(\phi_1,\phi_{2})_{\cal KG}=\int\!\! \goi\, W_{\!\mu}(\phi_1^\ast,\phi_2)\,{\rm d}\varSigma^\mu \label{KG}
		\end{equation}
where $W_\mu$ stands for Wronskian. For arbitrary pair $\phi_1$ and $\phi_2$ of complex solutions to the equation (\ref{dAlembert}) the divergence of Wronskian vanishes
		\begin{equation}
			\nabla^{\mu}W_\mu(\phi_1,\phi_2^\ast)=\nabla^{\mu}
			\begin{array}{|cc|}
				\phi_{1} &\phi_{2}^{*}\\
 				\nabla_{\mu}\phi_{1} &\nabla_{\mu}\phi_{2}^{*}
			\end{array}=
			\begin{array}{|cc|}
 				\phi_{1} &\phi_{2}^{*}\\
 				\nabla^{\mu}\nabla_{\mu}\phi_{1} &\nabla^{\mu}\nabla_{\mu}\phi_{2}^{*}
			\end{array}=0
		\end{equation}
This assures that the integral (\ref{KG}) is independent of the choice of the $\varSigma$ hypersurface. In particular, the integral (\ref{KG}) and all other quantities defined on this base are invariant under the gauge modifications. Coefficients $\ak$ are constant in time.
Normalisation (\ref{norma}) together with separation (\ref{product}) result in 
		\begin{equation}
W_{\!\mu}(\fchi,\fchi^\ast)=-\goi \label{Wronskianchi}
		\end{equation}
and
		\begin{equation}
\int\!\!{F(\sfx,\gon,k) F^{\ast}(\sfx,\gon,k')}\,{\rm d}\varSigma^\mu=\frac{k}{k-\goi\sqrt{-K}}\delta(k-k ')
		\end{equation}
To express the momentum (\ref{pedy1}) of the wave  moving in the direction $n^{i}$ we write
		\begin{eqnarray}
\pi(n_{i} K^{(i)}_M)
&=& \int\!\! {\rm d}\varSigma^{0} n_{i}K^{(i)\nu}_M {\cal T}_{{0}\nu}\nonumber\\
&=&\int\!\! {\rm d}\varSigma^{0} \partial_0 \gg n_{i} K^{(i)\nu}_M \partial_\nu \gg 
		\end{eqnarray}
On the strength of (\ref{dekompozycja}), (\ref{norma}), (\ref{wspolczynniki1}) and (\ref{Wronskianchi}) after laborious but straightforward calculations one obtains
		\begin{eqnarray}
\pi(n_{i} K^{(i)}_M)&=&
\ikk \ak\akprimast \int\!\! {\rm d}\varSigma^{0} \left[
(n_{i} K^{(i)\nu}_M\partial_\nu \uk)(\partial_0 \ukprimast)
+(n_{i} K^{(i)\nu}_M\partial_\nu \ukprimast)(\partial_0 \uk)\right]\nonumber \\
& =& \ik(\ak\akast \goi\, k W_{0}(\chik,\chikast))\nonumber \\
& =& \ik (k \ak\akast)
= \ik (k P_k)\label{spec} 
		\end{eqnarray}
The quantity $P_k=a_k\, a^*_k$ is constant in time and invariant under both: the gauge transformations of the reference system, and the unitary transformations of Fourier bases $\uu$. $P_k$ is an observable. It is the intrinsic property of the acoustic field  and, according to (\ref{spec}), can be regarded as the {\it momentum spectrum\/} or {\it hyperbolic momentum spectrum\/} of sound.

\section{Summary}
\label{sec:08}

The presented gauge-invariant formalism is the first formalism which essentially unifies both subhorizon and superhorizon regimes. There is no need to divide solutions into classes according to their length scale\footnote{Except for non vanishing curvature or cosmological constant.}.
This particular property is due to the fact that 
the perturbation variables $\delta\rho/\rho$ defined in the original spacetime form the massless scalar field $\gamma$ in the corresponding {\em acoustic spacetime}. Fourier coefficients $a_k$ are based on the Klein-Gordon scalar product (\ref{KG}), and therefore, they are gauge-independent and constant in time. Frequency distribution of the momentum is described by the hypersurface-independent spectrum $P_k$.  

The acoustic spacetime enable one to construct the energy momentum tensor of sound. The six-parameter isometry group acting in this space results in the conservation of momentum, hyperbolic momentum, and angular momentum of sound waves.
The momentum conservation (\ref{pedy1}) show that the physical nature of waves, whether ``inside'' or ``outside'' the particle horizon, is the same. The wave momentum does not change at the {\em horizon crossing\/}, consequently, the phase and group velocity also remains the same. 
The evidence of the momentum conservation testify that the expected {\em freezing\/} of waves larger than the horizon-size is rather a mathematical effect caused by neglecting some terms in differential equations. 

Although the total momentum density in the Robertson--Walker universe is expected to be small, the momenta of the constituent waves may be  arbitrary. There are no adequate physical conditions necessary for exciting and maintaining standing waves~\cite{66}. The generic solution to the perturbation equation is the running wave. The absence of time-like isometry results in the absence of the energy conservation. Acoustic waves may undergo  {amplification and backscattering}, due to the same mechanism which works in the gravitational waves theory~\cite{61,62}.

The perturbation spectrum depends on curvature. In the open universe scalar perturbations are composed of acoustic modes with real wavenumbers $0<k^2$ (principal series) and non-acoustic modes with imaginary wavenumbers $-1<k^2<0$ (supplementary series). Both series of modes may affect the temperature fluctuations, and particularly contribute to the lowest multipoles. The dispersion on curvature, dispersion on cosmological constant and the gravitational waves amplification affect the same range of scales. This may impede to separate these phenomena in the CMBR data. 

The lack of appropriate identification of additive constants of motion in cosmology was the essential obstacle to  extend the statistical physics theory to the expanding universe.
Removing this obstacle by reformulating the problem in terms of acoustic geometry
enables one to replicate the fluid statistical mechanics~\cite{67} to universes with Robertson--Walker symmetry. Moreover, for the radiation-dominated epoch the resulting acoustic space time is static, and consequently, in this particular but important case also the energy conservation is restored. The construction of statistical fluid cosmology is beyond the scope of this paper, but we appreciate its importance. 
The classical fluid fluctuations may  admix to the temperature spectrum, preventing us from knowing which of the observed fluctuations are credible tracers of the early quantum processes.


\def\text{\mbox}

\section{{\em Mathematica 7.0} code --- Appendix}
\label{sec:09}

\medskip
\leftline{(* Preliminary definitions *)}
\medskip

\noindent
$\text{ClearAll[{\sf ''}Global`*{\sf ''}]}$\\
$X=\{\eta ,\sfx ,\sfy ,\sfz \};$
$x=\text{Sequence}\text{@@}X;$\\
$\XX=\{\eta \_,\sfx \_,\sfy \_,\sfz \_\};$ 
$\xx=\text{Sequence}\text{@@}\XX;$
\medskip

\noindent
$Y=\text{Rest}[X];$
$y=\text{Sequence@@}Y;$\\
$\DD=\{p1,p2,p3\};$\\

$\DDD=\{p1\_,p2\_,p3\_\};$
$\ddd=\text{Sequence@@}\DDD;$\\
$\text{spacedif}=\text{Sequence@@Thread}[\{\text{Y},\DD\}];$
\medskip 

\noindent
$\text{zero}=\text{Sequence@@ConstantArray}[0,\text{Length}[Y]];$

\leftline{}
\leftline{(* Space metric tensor *)}
\smallskip

\noindent 
$
g=\text{DiagonalMatrix}[\{\gog[y],\gog[y],\gog[y]\}];
$

\smallskip
\leftline{}
\leftline{(* Friedman equations *)}
\smallskip

\noindent
$FR1=\frac{1}{3}\epsilon[\eta]==\frac{a'[\eta]^2}{a[\eta]^4}+\frac{K}{a[\eta]^2};$\\
$FR2=p[\eta]==-2\frac{{a''}[\eta]}{a[\eta]^3}+\frac{{a'}[\eta]^2}{a[\eta]^4}-\frac{K}{a[\eta]^2};$

\smallskip
\leftline{}
\leftline{(* Background dynamics *)}
\medskip

\noindent 
$a'[\eta\_]=a[\eta]^2H[\eta];$\\
$\epsilon'[\eta\_]=-3a[\eta]H[\eta](\epsilon[\eta]+p[\eta]);$\\
$H'[\eta\_]=-a[\eta]\left(H[\eta]^2+\frac16(\epsilon[\eta]+3p[\eta])\right);$
\medskip

\noindent 
\(\text{eps}=\text{Solve}[FR1,\epsilon[\eta]][\![1,1]\!];\)\\[.2ex]
\(\epsilon[\eta\_]=\text{Simplify}[\epsilon[\eta]/.\text{eps}];\)

\smallskip
\leftline{}
\leftline{(* generalised Lifshitz equations: $\lambda=\llambda$, $\mu=-\mmu$  *)}
\smallskip

\noindent
\({{\lambda ^{(2,\text{zero})}[\xx]=-2\frac{a'[\eta ]}{a[\eta]}
\peta
\lambda[x]-\frac{1}{3} (\text{Lap}[\lambda][x]+\text{Lap}[\mu
][x]);}}\\
{{\mu ^{(2,\text{zero}))}[\xx]=-\left(2+3 c[\eta ]^2\right) \frac{ a'[\eta ] }{a[\eta ]}\peta
\mu [x]+\frac{1}{3}\left(1+3
c[\eta ]^2\right)}}\)

\noindent~~~~~~~~~~~~~~~~~~~~~~~~~~~~~~~~~~~~~~~~~~~~~~~~~~~~~~~~~~~~~~~
 \(
(\text{Lap}[\lambda ][x]+\text{Lap}[\mu ][x]+3 K(\lambda [x]-\mu [x]));
\)

\smallskip
\leftline{}
\leftline{(* The density contrast *)}

\smallskip

\noindent
\({{\delta [{\xx}]}={ }{\frac{1 }{3 a[\eta ]^2\epsilon [\eta ]}\left( -(\text{Lap}[\lambda ][x]+\text{Lap}[\mu][x]+3 K( \lambda [x]+ \mu [x]))+
3\frac{a'[\eta ]}{a[\eta ]}
\peta
\mu [x]\right)}}\);

\smallskip
\leftline{}
\leftline{(* Derivative reccurence commands *)}
\smallskip

\noindent 
$\lambda ^{(\text{k$\_$Integer}\text{/;}k>2,\text{zero})}[{\xx}]:=\partial_{\{\eta,k-2\}}\lambda ^{(2,\text{zero})}[x]$\\[0.2ex]
$\mu ^{(\text{k$\_$Integer}\text{/;}k>2,\text{zero})}[{\xx}]:=\partial_{\{\eta,k-2\}}\mu ^{(2,\text{zero})}[x]$
\medskip

\noindent 
\(\lambda ^{(2,\ddd)}[{\xx}]:=\partial_{\text{spacedif}}\lambda ^{(2,\text{zero})}[x]\)\\[.2ex]
\(\mu ^{(2,\ddd)}[{\xx}]:=\partial_{\text{spacedif}}\mu ^{(2,\text{zero})}[x]\)

\smallskip 
\leftline{}
\leftline{(* Laplacian *)}
\smallskip

\noindent
$\text{Laplasjan}[x\_,g\_]:=\text{Module}[\{\},\text{dim}=\text{Length}[x];\\
\triangle_{\text{dim}}:=\frac{1}{\sqrt{\text{Det}[g]}}
\sum\limits_{i=1}\limits^{\text{dim}}\sum\limits_{k=1}^{\text{dim}}
\partial_{x[\![i]\!]}
\left(
\sqrt{\text{Det}[g]}\,\text{Inverse}[g][\![i, k]\!]\partial_{x[\![k]\!]}\#1\right)\&]
$

\medskip 

\noindent 
$
\text{Laplasjan}[Y,g]$;
$\text{Lap}[f\_][\xx]=\triangle_{\text{dim}}[f[x]]//\text{Simplify};
$

\smallskip 
\leftline{}
\leftline{(* Canonical variable *)}
\smallskip

\noindent 
$\text{Clear}[H]$\\
$c[\eta\_]=\sqrt{\frac{p'[\eta]}{\epsilon'[\eta]}}$; 
$\goa[\eta\_]={a[\eta]}\sqrt{\frac{p[\eta]+\epsilon[\eta]}{3c[\eta]{H^2[\eta]}}}$;
$b[\eta\_]=\frac{1}{a[\eta]}\frac{\epsilon'[\eta]}{\epsilon[\eta]}$;
\medskip

\noindent 
$\gamma[\xx]=a[\eta] H[\eta]^2 \peta\frac{\delta[x]}{b[\eta]}//\text{Simplify};$

\leftline{}
\leftline{(* The wave equation *)}
\smallskip

\noindent 
$\text{wave}=\petaeta{\gamma[x]}+
\left[2\frac{\goa'[\eta]}{\goa[\eta]}-\frac{c'[\eta]}{c[\eta]}
\right]\peta{\gamma[x]}-c[\eta]^2\text{Lap} [\gamma][x]
==0;$
\medskip

\noindent 
$\text{Timing}[\text{Reduce}[\text{wave}/.\text{eps}]]$


\begin{thebibliography}{00}

	\bibitem{1}
E. Lifshitz, {\em J. Phys.} {\bf 10} (1946) 116.

	\bibitem{2}
E. M. Lifshitz and M. Khalatnikov, {\em Adv. Phys.} {\bf 12} (1963) 185.

	\bibitem{3}
L. D. Landau and  E. M. Lifshitz, {\em  Course of Theoretical Physics}, {\em Volume 2: The Classical Theory of Fields}, Butterworth-Heinemann, Elsevier (2000).

\bibitem{4}
J. Bardeen, {\em Phys. Rev. D} {\bf  22} (1980) 1882.

	\bibitem{5}
R. Brandenberger, R. Kahn and W. Press, {\em Phys. Rev. D} {\bf 28} (1983) 1809.

	\bibitem{6}
J. Katz, J. Bicak and D. Lynden-Bell, {\em Phys. Rev. D}  {\bf 55} (1997) 5957.

	\bibitem{7}
N. Deruelle, J. Katz and J. P. Uzan, {\em Class. Quantum Grav.} {\bf 14} (1997) 421. 

	\bibitem{8}
N. Deruelle and J. P. Uzan, {\em Int. J. Theor. Phys.} {\bf 36} (1997) 2461.

	\bibitem{9}
D. H. Lyth and M. Mukherjee, {\em Phys. Rev. D}  {\bf 38} (1988) 485.

	\bibitem{10}
V. Mukhanov, H. Feldman and R. Brandenberger, {\em Physics Report\/} {\bf 215} (1992) 
203.

	\bibitem{11}
S. Hawking, {\em Astrophys. J.} {\bf 145} (1966) 544.

	\bibitem{12}
D. W. Olson,  {\em Phys. Rev. D\/} {\bf 14} (1976) 327.

	\bibitem{13}
A. Woszczyna and A. Kulak,  {\em Class. Quantum Grav.} {\bf 6} (1989) 1665.

	\bibitem{14}
G. F. R. Ellis and M. Bruni, {\em Phys. Rev. D\/} {\bf 40} (1989) 1804.

	\bibitem{15}
G. F. R. Ellis, M Bruni and J. Hwang, {\em Phys. Rev. D\/} {\bf 42} (1990) 1035.

	\bibitem{16}
R. K. Sachs and A. M. Wolfe, {\em Astrophys. J.} {\bf 147} (1967) 73.

	\bibitem{17}
G. B. Field and L. C. Shepley, {\em Astrophys. J.} {\bf 1} (1968) 309.

	\bibitem{18}
V. N. Lukash,  {\em JETP} {\bf 79} (1980) 1601.

	\bibitem{19}
G. V. Chibisov and F. Mukhanov, {\em Mon. Not. Roy. Astron. Soc.} {\bf 200} (1982) 
535.

	\bibitem{20}
S. D. Brechet, M. P. Hobson and A. N. Lasenby [http://lanl.arxiv.org/abs/0909.5384].

	\bibitem{21}
G. B. Whitham,  {\em J. Fluid. Mech.}  {\bf 12} (1962) 135.

	\bibitem{22}
G. B. Whitham, {\em J.\ Fluid.\  Mech.}  {\bf 22} (1965) 273.

	\bibitem{23}
C. J. R. Garrett, {\em Proc.\ Roy.\ Soc. A\/} {\bf 299} (1967) 26.

	\bibitem{24}
J. Lighthill,  {\em Waves in Fluids}, Cambridge University Press, Cambridge (1978).


	\bibitem{25}
D. G. Andrews and M. E. McIntyre, {\em J.\  Fluid.\  Mech.} {\bf 89} (1978) 609.

	\bibitem{26}
D. G. Andrews and M. E. McIntyre, {\em J.\  Fluid.\  Mech.} {\bf 89} (1978) 647.

	\bibitem{27}
M. K. Meyers, {\em J.\  Fluid.\  Mech.} {226} (1991) 383.

	\bibitem{28}
R. H. Cantrell and R. W. Hart,  {\em J.\ Acoustical Soc.\  Amer.} {36} (1964) 697.

	\bibitem{29}
O. S. Ryshov and  G. M. Shefter,  {\em Prik.\ Mat.\ Mek.} {\bf 26} (1962) 854.

	\bibitem{30}
O. S. Ryshov and  G. M. Shefter,  {\em J.\ Appl.\ Math.\ Mech.} {\bf 26} (1962) 1293.

	\bibitem{31}
 C. L. Morfey, {\em J.\ Sound and Vib.} {\bf 14} (1971) 159.

	\bibitem{32}
M. K.  Meyers, {\em J.\ Sound and Vib.} {\bf 109} (1986) 277.

	\bibitem{33}
M. E. McIntyre, {\em J.\  Fluid.\  Mech.} {\bf 106} (1981) 331.

	\bibitem{34}
W. Unruh, {\em Phys. Rev. Lett.} {\bf 46} (1981) 1351.

	\bibitem{35}
W. Unruh,  {\em Phys. Rev. D\/} {\bf 51} (1995) 2827.

	\bibitem{36}
M. Visser, {\em Class. Quantum Grav.} {\bf 15} (1998) 1767.

	\bibitem{37}
M. Stone, {\em Phys. Rev. E\/} {\bf 62} (2000) 1341.
 
	\bibitem{38}
R. Peierls, {\em Surprises in Theoretical Physics}, Princeton  (1979).

	\bibitem{39}
N. Jacobson, {\em Lie algebras}, Dover Publications (1979).

	\bibitem{40}
 Z. A. Golda and A. Woszczyna, {\em Class. Quantum Grav.} {\bf 18} (2001) 543.

	\bibitem{41}
 Z. A. Golda and A. Woszczyna, {\em Phys. Lett. A\/} {\bf 310} (2003) 357.

	\bibitem{42}
 Z. A. Golda and A. Woszczyna, {\em Class. Quantum Grav.} {\bf 20} (2003) 277.

	\bibitem{43}
N. D. Birrell and P. C. W. Davies,  {\em Quantum fields on curved space}, 
Cambridge University Press,  Cambridge (1982).

	\bibitem{44}
J. M. Stewart, {\em Class. Quantum Grav.} {\bf 7} (1990) 1169.

	\bibitem{45}
W. Czaja,  PhD Thesis, Jagiellonian University, Krak\'ow (2009).

	\bibitem{46}
M. A. M. Santiago and A. N. Vaidya, {\em J. Phys.  A: Math. Gen.} {\bf 9} (1976) 897.

	\bibitem{47}
N. J. Korevaar, R. Kusner, W. H. Meeks and B. Solomon, {\em  American 
Journal of Mathematics\/} {\bf 114} (1992) 1.

	\bibitem{48}
A. Staruszkiewicz, {\em Acta Phys. Polonica B\/} {\bf 26} (1995) 1275.

	\bibitem{49}
 M. Suffczy\'nski, {\em Electrodynamics}, in polish, PWN, Warszawa (1965).

	\bibitem{50}
J. D. Jackson, {\em Classical Electrodynamics},  third edition, John Wiley \& Sons, Inc. (1999). 
 
	\bibitem{51} 
A. Staruszkiewicz, {\em Lectures on classical electrodynamics},  unpublished, 
Jagiellonian University (1990--1996).

	\bibitem{52}
 M. A. Alonso,  G. S. Pogosyan, and K. B. Wolf, {\em J. Math. Phys.} {\bf 43} (2002) 5857.

	\bibitem{53}
I. S. Shapiro, {\em Phys. Lett.} {\bf 1} (1962) 253.

	\bibitem{54}
V. G. Kadyshevsky, R. M. Mir-Kasimov and N. B. Skachkov, {\em Problems 
Elementary Particle At. Nucl. Phys.} {\bf 2} (1972) 635.

	\bibitem{55}
A. M. Yaglom, in {\em Proceedings of the Fourth Berkeley Symposium}, Volume II, 
edited by J.~Neyman, University of California Press, Berkeley (1961).

	\bibitem{56}
D. H. Lyth and A. Woszczyna, {\em Phys. Rev. D\/} {\bf 52} (1995) 3338.

	\bibitem{57}
A. Staruszkiewicz, {\em Acta Phys. Polonica B} {\bf 23} (1992) 591.

	\bibitem{58}
T. O. Sherman, {\em Trans. Am. Math. Soc.} {\bf 209} (1975) 1.

	\bibitem{59}
 I. P. Volobuyev, {\em Teor. Mat. Fiz.} {\bf 45} (1980) 421.

	\bibitem{60}
M. A. Alonso, G. S. Pogosyan and K. B. Wolf, {\em J. Math. Phys.} {\bf 44} (2003) 1472.

	\bibitem{61}
 F. L. Abbott and D. D. Harari,  {\em Nucl. Phys.} {\bf 264} (1986) 487.

	\bibitem{62}
 L. P. Grishchuk, in {\em Quantum Fluctuations: Les Houches}, Session LXIII,  
Elsevier (1995), p. 541.

	\bibitem{63}
M. Goetz, {\em Mon. Not. Roy. Astron. Soc.} {\bf 295} (1998) 873.

	\bibitem{64}
S. Weinberg, {\em Phys. Rev. D\/} {\bf 67} (2003) 123504.

	\bibitem{65}
 L. P. Grishchuk [arXiv:gr-qc/9511074v1].

	\bibitem{66}
R. P. Feynman, R. B. Leighton, M. Sands, {\em The Feynman Lectures on Physics}, vol. 1, part 2,  Addison --- Wesley (1963).

	\bibitem{67}
V. I. Kalikmanov,  {\em Statistical Physics of Fluids: Basic Concepts and Applications},
Springer (2001).

\end{thebibliography}
\end{document}